\documentstyle[a4wide,12pt,epsf]{article}
\begin{document}
\begin{flushright}
\baselineskip=12pt
{SUSX-TH/01-006}\\
{RHCPP01-01T}\\
{hep-ph/01mmdd}\\
{March 2001}
\end{flushright}

\begin{center}
{\LARGE \bf DARK MATTER CONSTRAINTS IN HETEROTIC M-THEORY WITH FIVE-BRANE DOMINANCE \\}
\vglue 0.35cm
{D.BAILIN$^{\clubsuit}$ \footnote
{D.Bailin@sussex.ac.uk}, G. V. KRANIOTIS$^{\spadesuit}$ \footnote
 {G.Kraniotis@rhbnc.ac.uk} and A. LOVE$^{\spadesuit}$ \\}
	{$\clubsuit$ \it  Centre for Theoretical Physics, \\}
{\it University of Sussex,\\}
{\it Brighton BN1 9QJ, U.K. \\}
{$\spadesuit$ \it  Centre for Particle Physics , \\}
{\it Royal Holloway and Bedford New College, \\}
{\it  University of London,Egham, \\}
{\it Surrey TW20-0EX, U.K. \\}
\baselineskip=12pt

\vglue 0.25cm
ABSTRACT
\end{center}

{\rightskip=3pc
\leftskip=3pc
\noindent
\baselineskip=20pt
The phenomenological implications of the M-theory limit 
in which supersymmetry is broken by the auxiliary fields of five-brane 
moduli is investigated.
 Assuming that the lightest neutralino 
provides the dark matter in the universe, constraints on the 
sparticle spectrum are obtained. Direct detection rates for dark matter are 
estimated.}

\vfill\eject
\setcounter{page}{1}
\pagestyle{plain}
\baselineskip=14pt
	
One of the most interesting developments in M-theory 
\cite{HORWIT,witten,ANTO,NANO,LIS,NILLES,DUDAS,STEVE,CHOI,LUKAS} 
model building 
is that new non-perturbative tools have been developed which 
allow the construction of realistic three generation models \cite{BURT}. 
In particular the inclusion of five-brane moduli $Z_n$, (which do not have 
a weakly coupled string theory counterpart) besides the 
metric moduli $T,S$ in the effective 
action leads to new types of $E_8 \times E_8$ symmetry breaking patterns  
as well as to novel gauge and K$\rm {\ddot a}$hler threshold 
corrections. As a result the soft-supersymmetry breaking terms 
differ substantially from the weakly coupled string.

The phenomenological implications of the effective action of M-theory 
with the standard embedding of the spin connection into the gauge fields
have been investigated in \cite{BKL,KINE,CARLOS,CASA,LI}. 
Some phenomenological 
implications of non-standard embeddings in M-theory with and without 
five-branes have been 
studied in \cite{TATSUO,MUNOZ}.

In a previous letter, we investigated the supersymmetric 
particle spectrum in the interesting case 
when the auxiliary fields associated with the five-branes dominate  
those associated with metric moduli  
($F^{Z_n} \gg F^S,F^T$), including 
the constraint of radiative electroweak symmetry-breaking \cite{BKLbrane}.
It is the purpose of this letter, to extend the calculation and take 
into account cosmological constraints on the relic abundance of the  
neutralino assuming 
it provides the dark matter of the universe
in the region of the parameter space  in which 
it is the lightest supersymmetric particle. As we 
shall see in what follows 
these constraints are quite restrictive.

The soft supersymmetry-breaking 
terms are determined by the following functions of the effective 
supergravity theory \cite{BURT,MUNOZ}:
\begin{eqnarray}
K&=&-{\rm ln}(S+\bar{S})-3{\rm ln}(T+\bar{T})+K_5+\frac{3}{T+\bar{T}}(
1+\frac{1}{3}e_O)H_{pq}C_O^p{\bar{C}^q_O}
, \nonumber \\
f_{O}&=&S+B_O T, \;\;f_{H}=S+B_H T, \nonumber \\
W_O&=&d_{pqr}C^p_O C^q_O C^r_O
\label{mfunc}
\end{eqnarray}
where $K$ is the K$\rm{\ddot{a}}$hler potential, $W_O$ the 
observable sector perturbative superpotential, 
$C^p_O$ are observable sector matter fields
and 
$f_{O}, f_{H}$ are the gauge kinetic functions for the observable 
and hidden sector gauge groups respectively.
$K_5$ is the K$\rm{\ddot a}$hler potential for the five-brane 
moduli $Z_n$ and $H_{pq}$ is some $T$-independent metric. 
Also
\begin{equation}
e_O=b_O \frac{T+\bar{T}}{S+\bar{S}},\;\;
e_H=b_H\frac{T+\bar{T}}{S+\bar{S}}
\end{equation}
and the coefficients 
$b_{O,H},B_{O,H}$ are given in terms of the 
instanton numbers $\beta_{O,H}$ and the five 
brane charges $\beta_n$ by the following expressions
\begin{eqnarray}
b_O &=&\beta_O+\sum_{n=1}^{N}(1-z_n)^2 \beta_n \nonumber \\
B_O &=&\beta_O+\sum_{n=1}^{N}(1-Z_n)^2 \beta_n \nonumber \\
B_H &=&\beta_H+\sum_{n=1}^{N}(Z_n)^2 \beta_n \nonumber \\
b_H &=&\beta_H+\sum_{n=1}^{N}z_n^2 \beta_n
\end{eqnarray}
and the five-brane moduli are denoted by $Z_n$ whose 
${\rm{Re}} Z_n \equiv z_n=\frac{x_n}{\pi \rho} \in (0,1)$ 
are the five-brane positions in the normalized orbifold 
coordinates.
Since a Calabi-Yau manifold is compact, the net magnetic charge 
due to orbifold planes and 5-branes is zero. Consequently the 
following cohomology condition is satisfied
\begin{equation}
\beta_O+\sum_{n=1}^{N}\beta_n+\beta_H=0
\end{equation}
$S,T$ are the dilaton and 
Calabi-Yau moduli fields and $C^{p}$ charged matter fields.
The superpotential
and the gauge kinetic functions are exact up to non-perturbative effects.

Given eqs(\ref{mfunc}) one can determine 
\cite{BURT,MUNOZ} the soft supersymmetry breaking terms
for the 
observable sector 
gaugino masses $M_{1/2}$, scalar masses $m_0$ and trilinear scalar 
couplings $A$
as functions of the auxiliary fields $F^S$,$F^T$, $F^{n}$ of the moduli 
$S,T$ fields and five-brane moduli $Z_n$ respectively.
\begin{eqnarray}
M_{1/2}&=&
\frac{1}{(S+\bar{S})(1+\frac{B_O T+\bar{B}_O \bar{T}}{S+\bar{S}})}
(F^S+F^T B_O+T F^n \partial_n B_O) \nonumber \\
m_0^2&=&V_0+m_{3/2}^2-\frac{1}{(3+e_O)^2}\Bigr[
e_O (6+e_O) \frac{|F^S|^2}{(S+\bar{S})^2} \nonumber \\
&+& 3(3+2e_O)\frac{|F^T|^2}{(T+\bar{T})^2}-
\frac{6 e_O}{(S+\bar{S})(T+\bar{T})} Re F^S {\bar{F}^{\bar{T}}} \nonumber \\
&+& \Bigl(\frac{e_O}{b_O}(3+e_O)\partial_n \partial_{\bar{m}} 
b_O-\frac{e_O^2}{b_O^2}\partial_n b_O 
\partial_{\bar{m}} b_O\Bigl) F^n \bar{F}^{\bar{m}} \nonumber \\
&-&\frac{6 e_O}{b_O}\frac{\partial_{\bar{n}} b_O}{S+\bar{S}}
Re F^S \bar{F}^{\bar{n}}+\frac{6 e_O}{b_O}
\frac{\partial_{\bar{n}} b_O}{T+\bar{T}} Re F^T \bar{F}^{\bar{n}}
\Bigr], \nonumber \\
A&=&-\frac{1}{3+e_O}\Bigl\{\frac{F^S (3-2 e_O)}{S+\bar{S}}+
\frac{3 e_O F^T}{T+\bar{T}} \nonumber \\
&+& F^n \Bigl(3\frac{e_O}{b_O}\partial_n b_O-(3+e_O)\partial_n K_5 \Bigr)
\Bigr\}
\label{GAUG}
\end{eqnarray}
where $\partial_n \equiv \frac{\partial}{\partial Z_n}$.
The  bilinear $B$-parameter  associated with a non-perturbatively generated 
$\mu$ term in the superpotential 
is given by \cite{MUNOZ}:
\begin{eqnarray}
B_{\mu}&=& \frac{F^S (e_O-3)}{(3+e_O)(S+\bar{S})}-
\frac{3 (e_O+1) F^T}{(T+\bar{T})(3+e_O)} \nonumber \\
&+&\frac{1}{3+e_O}\Bigl[(3+e_O)F^n\partial _n K_5-2F^n 
\frac{e_O}{b_O}\partial_n b_O\Bigr]-m_{3/2} \nonumber \\
\label{beta}
\end{eqnarray}
From now on we assume that only one five-brane contributes to 
supersymmetry-breaking \footnote{We assume negligible 
$CP$-violating phases in the soft terms.}.
Then the  auxiliary fields are given by \cite{MUNOZ,IBA:Spain}
\begin{eqnarray}
F^1&=&\sqrt{3} m_{3/2} C (\partial_1 \partial_{\bar{1}} K_5)^{-1/2} 
\sin\theta_1 \nonumber \\
F^S &=& \sqrt{3} m_{3/2} C (S+\bar{S}) \sin\theta \cos\theta_1 \nonumber \\
F^T &=& m_{3/2} C (T+\bar{T}) \cos\theta \cos\theta_1 
\end{eqnarray}
The goldstino angles are denoted by $\theta,\theta_1$, $m_{3/2}$ is 
the gravitino mass and $C^2=1+\frac{V_0}{3 m_{3/2}^2}$ 
with $V_0$ the tree level vacuum 
energy density. The five-brane dominated supersymmetry-breaking 
scenario corresponds to $\theta_1=\frac{\pi}{2}$, i.e 
$F^T,F^S=0$, and we take    
the 
five brane which contributes to supersymmetry breaking to be located 
at $z_1=1/2$ in the orbifold interval. We also set $C=1$ in the 
above expressions assuming zero cosmological constant.

The resulting supersymmetric particle spectrum for 
a single five-brane present with 
$z_1=\frac{1}{2}$ 
has been investigated in \cite{BKLbrane}.
Our parameters are, $e_O$, $\partial_1 \partial_{\bar{1}} K_5$,
$\partial_1 K_5$, 
$m_{3/2}$, $sign\; \mu$ (which is not determined by the radiative 
electroweak symmetry breaking constraint), where 
$\mu$ is the Higgs mixing parameter in the 
low energy superpotential. 
The ratio of the two Higgs vacuum 
expectation values $\tan\beta=\frac{<H_2^0>}{<H_1^0>}$ is also 
a free parameter if we leave  
$B$ to be  determined 
by  the minimization of the one-loop Higgs effective potential. 
If $B$ instead is given by
(\ref{beta}), one determines the value of $\tan\beta$. 
For this purpose  we take $\mu$ independent of $T$ and $S$
because of our lack of knowledge of 
$\mu$ in $M$-theory.
We treat $e_O$ as a free parameter as the problem of 
stabilizing the dilaton and other moduli has not yet been solved, 
although there has been an interesting work in this area \cite{KOREA}.

The instanton numbers are model dependent. In this paper 
we  choose to work with the interesting example \cite{MUNOZ} with 
$\beta_O=-2$ and $\beta_1=1$ which implies $b_O=-7/4$. 
This implies that $b_H=5/4$ and allow us to study the region of 
parameter space with
$-1<e_O \leq 0$, which is not accessible in strongly coupled M-theory 
scenarios with standard embedding.
We also choose $\partial_1 K_5=\partial_1 \partial_{\bar{1}}K_5=1$.
In an earlier paper \cite{BKLbrane} we have  investigated 
deviations from these values 
and have verified the robustness of our results. 

We  use the following experimental bounds from unsuccessful 
searches at LEP and Tevatron for supersymmetric particles 
\cite{EUC}. 
We require the lightest 
chargino $M_{\chi^+_1}\geq  90$ GeV, 
and the lightest Higgs, $m_{h_0} \geq 83$ GeV. A lower 
limit on the mass of the lightest stop  $m_{\tilde{t}_2} >86$ GeV, 
from $\tilde{t}_2 \rightarrow c 
\chi_1^0$ decay  in  D$0$ is imposed. The stau 
mass eigenstate ($\tilde{\tau}$) should be heavier 
than 81 GeV from LEP2 results.

The soft masses start running from 
a mass $R_{11}^{-1}\sim 7.5 \times 10^{15}$ GeV with 
$R_{11}$ the extra $M$-theory dimension. 
Then using (\ref{GAUG}),(\ref{beta}) as boundary 
conditions for the soft terms,
one evolves the renormalization 
group equations down to the weak scale and determines 
the sparticle 
spectrum compatible with the constraints of correct electroweak symmetry 
breaking and the above experimental 
constraints on the sparticle 
spectrum.  

Electroweak symmetry breaking is characterized by the extrema equations 
\begin{eqnarray}
\frac{1}{2}M_Z^2&=&\frac{\bar{m}^2_{H_1}-\bar{m}^2_{H_2}\tan^2 \beta}
{\tan^2 \beta -1}-\mu^2 \nonumber \\
-B\mu&=&\frac{1}{2}(\bar{m}^2_{H_1}+\bar{m}^{2}_{H_2}+2\mu^2)\sin 2\beta
\end{eqnarray}
where 
\begin{equation}
\bar{m}^2_{H_1,H_2}\equiv m^2_{H_1,H_2}+\frac{\partial \Delta V}{
\partial {v^2_{1,2}}}
\end{equation}
and $\Delta V=(64 \pi^2)^{-1} {\rm {STr}} M^4[ln (M^2/Q^2)-\frac{3}{2}]$ 
is the 
one loop contribution to the Higgs effective potential. We include
contributions only from the third generation of particles and sparticles.

Since $\mu^2 \gg M_Z^2$ for most of the allowed
region of the parameter space \cite{NATH}, the following 
approximate relationships 
hold at the 
electroweak scale for the 
masses of neutralinos and charginos, which of 
course depend on the details of 
electroweak symmetry breaking. 
\begin{eqnarray}
m_{\chi_1^{\pm}}\sim m_{\chi_2^0}\sim 2 m_{\chi_1^0} \nonumber \\
m_{\chi_{3,4}^{0}}\sim m_{\chi_2^{\pm}}\sim |\mu|
\label{neutralinos}
\end{eqnarray}
In (\ref{neutralinos}) $m_{\chi_{1,2}^{\pm}}$ are the chargino mass 
eigenstates and $m_{\chi_{i}^{0}},i=1\ldots 4$ are the four neutralino mass 
eigenstates with $i=1$ 
denoting the lightest 
neutralino. The former arise after diagonalization of the 
mass matrix. 
\begin{equation}
M_{ch}=\left(\begin{array}{cc}\\
M_2 & \sqrt{2} m_W \sin\beta \\
m_W \cos\beta & -\mu 
\end{array}\right)
\end{equation}
where $M_2$ denotes the weak gaugino mass and $M_1$ will denote the $U(1)_Y$ 
gaugino (Bino) mass.
The stau mass matrix is given by the expression 

\begin{equation}
{\cal M}_{\tau}^2=\left(\begin{array}{cc} \\
{\cal M}_{11}^2 & m_{\tilde{\tau}}(A_{\tau}+\mu \tan\beta) \\
m_{\tilde{\tau}}(A_{\tau}+\mu \tan\beta) & {\cal M}_{22}^2
\end{array}\right)
\end{equation}
where ${\cal M}_{11}=m^2_L+
m^2_{\tau}-
\frac{1}{2}(2M_W^2-M_Z^2)\cos 2 \beta$ and ${\cal M}_{22}=
m^2_E+m^2_{\tau}+(M_W^2-M_Z^2)\cos2\beta$.
where $m^2_L, \; m^2_E$ refer to scalar soft masses for lepton doublet, 
singlet respectively.


As has been first noted in \cite{MUNOZ}, in the case when only the 
five-branes contribute to supersymmetry-breaking the ratio of 
scalar masses to gaugino mass, $m_0/|M_{1/2}|>1$ for $e_O>-0.65$.
This is quite interesting since scalar masses larger than gaugino masses 
are not easy to obtain in the weakly-coupled heterotic string or 
M-theory compactification with standard embedding. 
As we shall see cosmological constraints become  important in this 
region of the parameter space.
In the case of non-standard 
embeddings without five-branes there is small region 
of the parameter space where is possible 
to have  $m_0/|M_{1/2}|>1$.

Assuming $R$-parity conservation 
the LSP is stable,  and consequently if it is neutral 
can provide a good dark matter candidate. 
We assume that the dark matter is in the form of neutralinos. 
The lightest neutralino  is a linear combination of the 
superpartners of the 
photon, $Z^0$ and neutral-Higgs bosons,
\begin{equation}
\chi_1^0=N_{11} \tilde{B}+N_{12}\tilde{W}^3+N_{13}\tilde{H}_1^0+N_{14}\tilde{H}_2^0
\end{equation}

The neutralino $4\times 4$ mass matrix can be written as
$$\left(\begin{array}{cccc} \\
M_1 & 0 & -M_Z A_{11} & M_Z A_{21} \\
0  & M_2 & M_Z A_{12} &-M_Z A_{22} \\
-M_Z A_{11} & M_Z A_{12} & 0 & \mu \\
M_Z A_{21} & -M_Z A_{22} & \mu & 0
\end{array}\right)$$ 
with 
$$\left(\begin{array}{cc} \\
 A_{11} & A_{12} \\
A_{21} & A_{22}\end{array}\right)= 
\left(\begin{array}{cc} \\
\sin\theta_{W} \cos\beta & \cos \theta_W \cos\beta \\
\sin\theta_W \sin\beta & \cos\theta_W \sin\beta 
\end{array}\right)$$

When the observational data on temperature fluctuations, 
type Ia supernovae, and gravitational lensing are combined with 
popular cosmological models, the dark matter relic abundance ($\Omega_{LSP}$) 
typically satisfies \cite{PERL}
\begin{equation}
0.1\leq \Omega_{LSP} h^2 \leq 0.4
\label{COSMO}
\end{equation}
where $h$ is the reduced Hubble constant.

We calculated the relic abundance of the lightest neutralino 
in the scenarios we have considered using standard 
techniques \cite{MARK}.
When these results are confronted with the (model-dependent) 
bounds (\ref{COSMO}) derived from the observational data 
further constraints on the parameters $m_{3/2},\tan\beta,\mu,e_O$ are 
obtained and these give new constraints on the sparticle spectrum.

\begin{figure}
\epsfxsize=6in
\epsfysize=8.5in
\epsffile{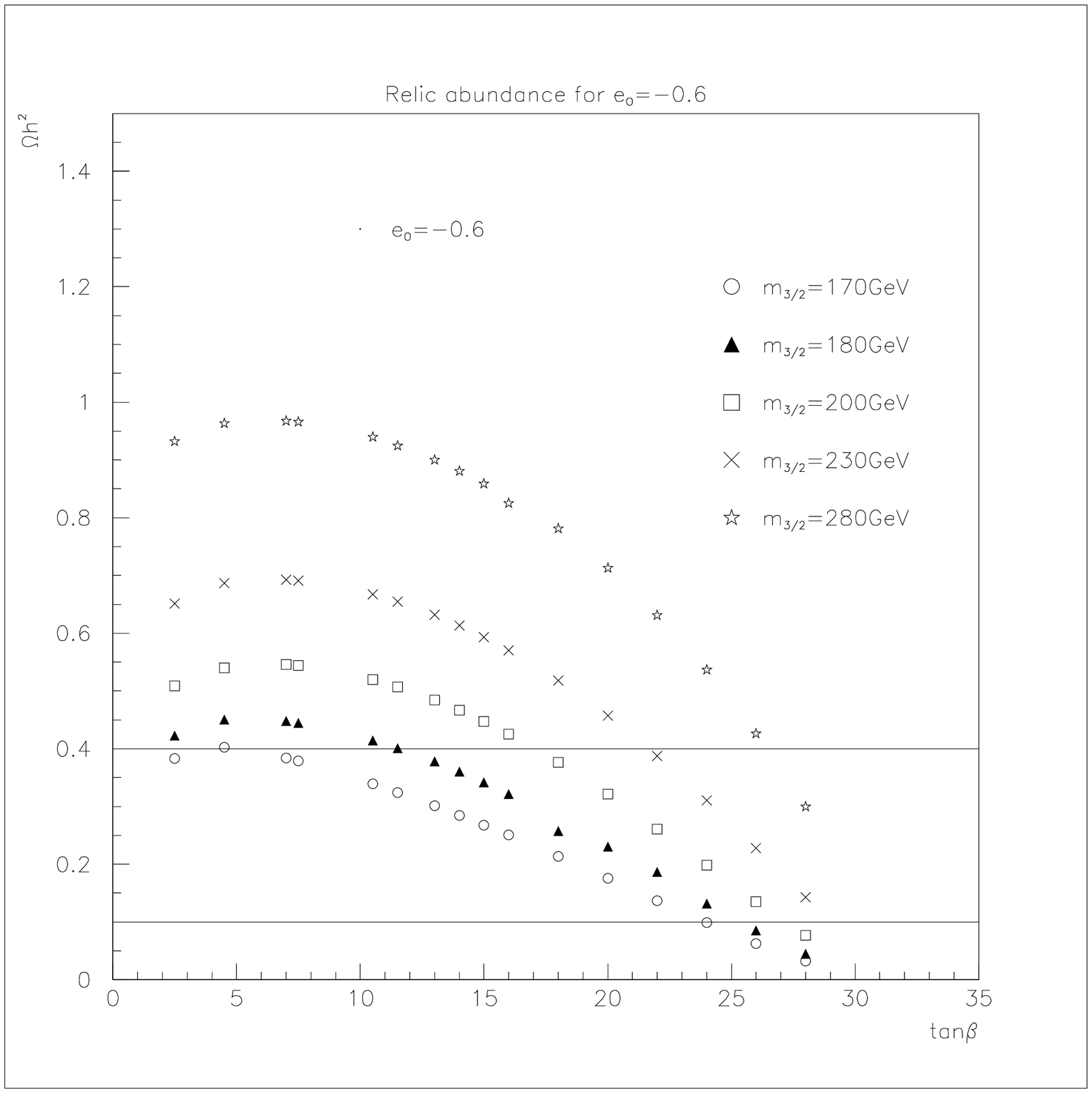}
\caption{Relic abundance of the LSP versus $\tan\beta$ for various values of 
the gravitino mass and $e_O=-0.6,\mu<0$. We also exhibit the upper and lower 
cosmological bounds on the LSP relic abundance}
\label{fige06}
\end{figure}

\begin{figure}
\epsfxsize=6.5in
\epsfysize=8.5in
\epsffile{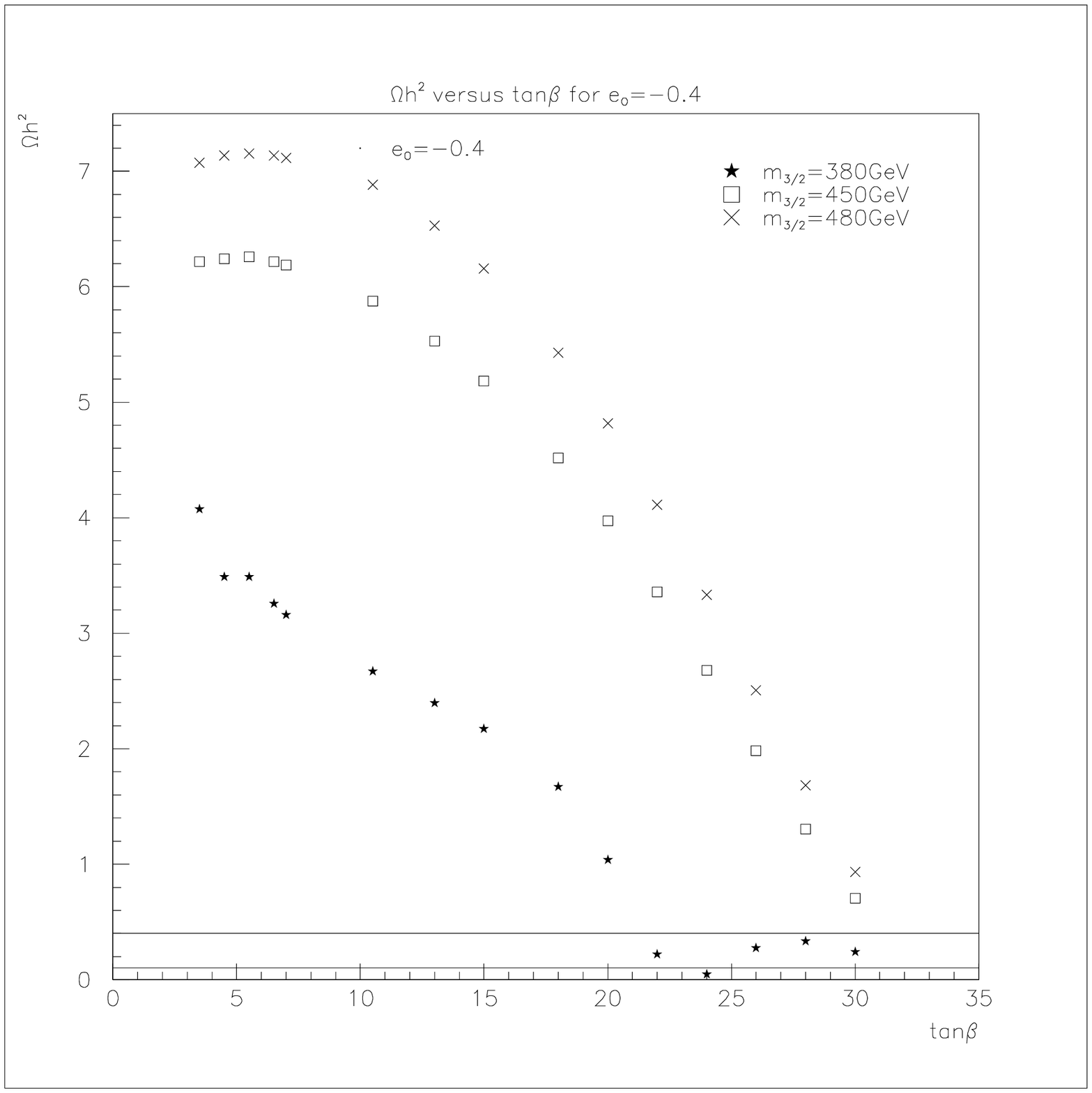}
\caption{Relic abundance of LSP vs $\tan\beta$ for $e_O=-0.4,\partial_1 K_5=
\partial_1 \partial_{\bar{1}} K_5=1,m_{3/2}=380GeV,450GeV,480GeV,\mu<0.$}
\label{fige04}
\end{figure}

\begin{figure}
\epsfxsize=6.8in
\epsfysize=8.5in
\epsffile{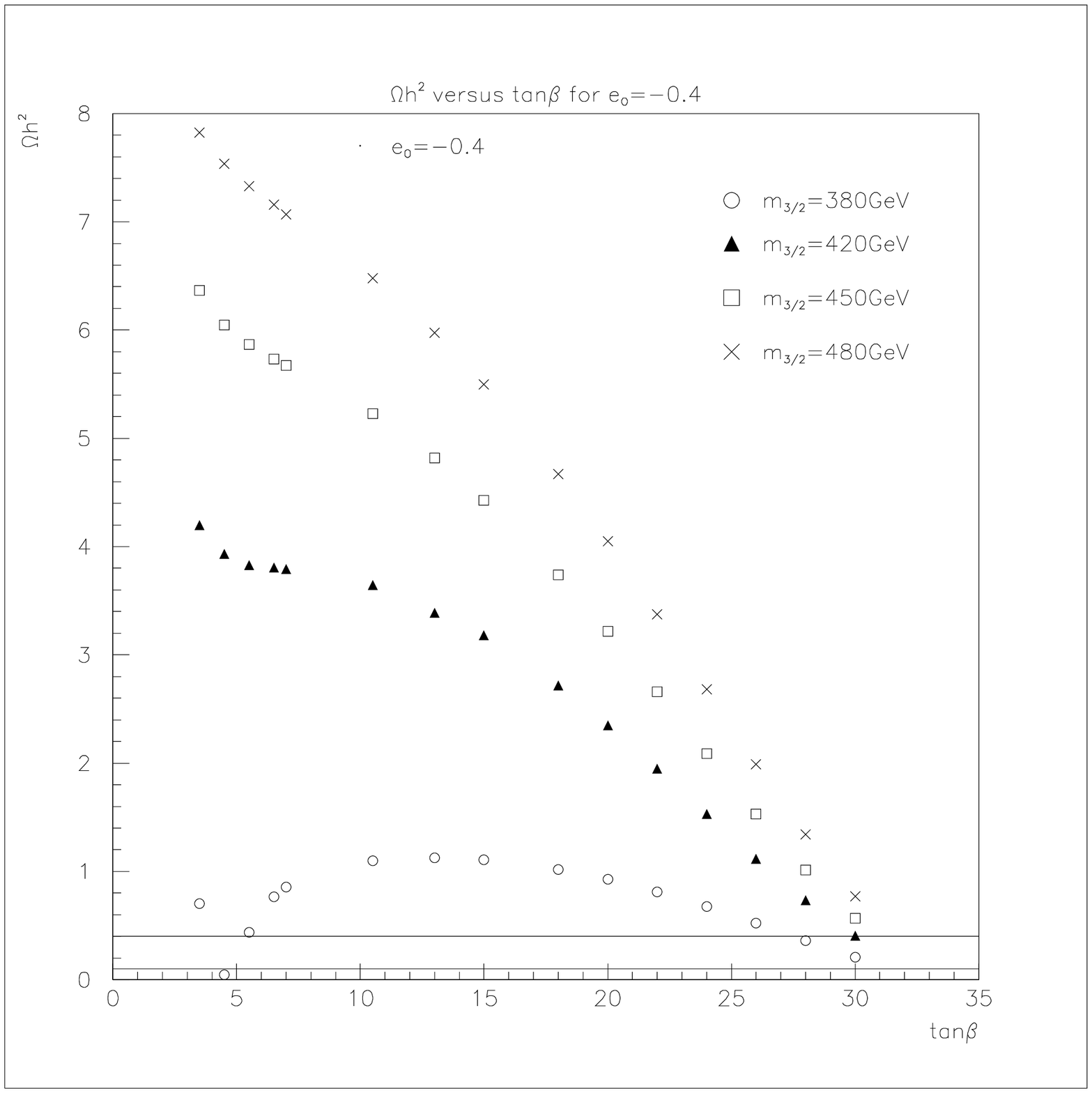}
\caption{Relic abundance of LSP vs $\tan\beta$ for $e_O=-0.4,\partial_1 K_5=
\partial_1 \partial_{\bar{1}} K_5=1,m_{3/2}=380GeV, 420 GeV, 450GeV,480GeV \mu>0$.}
\label{fige041}
\end{figure}

\begin{figure}
\epsfxsize=6in
\epsfysize=8.5in
\epsffile{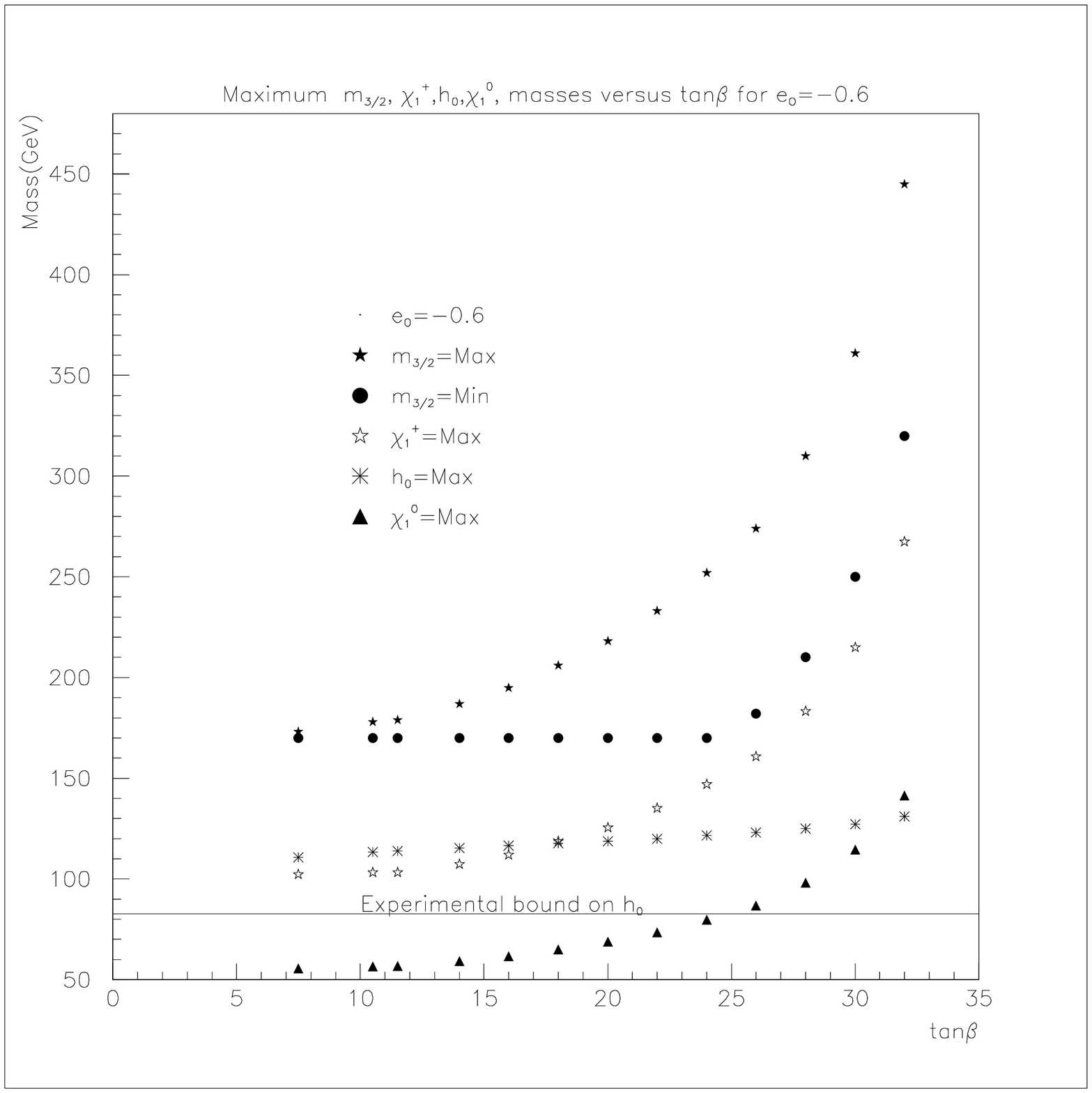}
\caption{Maximum gravitino, lightest chargino and lightest Higgs masses (imposed by 
$\Omega h^2 \leq 0.4$) versus $\tan\beta$ for $e_O=-0.6, \;\;\;\mu<0$.}
\label{grav}
\end{figure}

\begin{figure}
\epsfxsize=6in
\epsfysize=8.5in
\epsffile{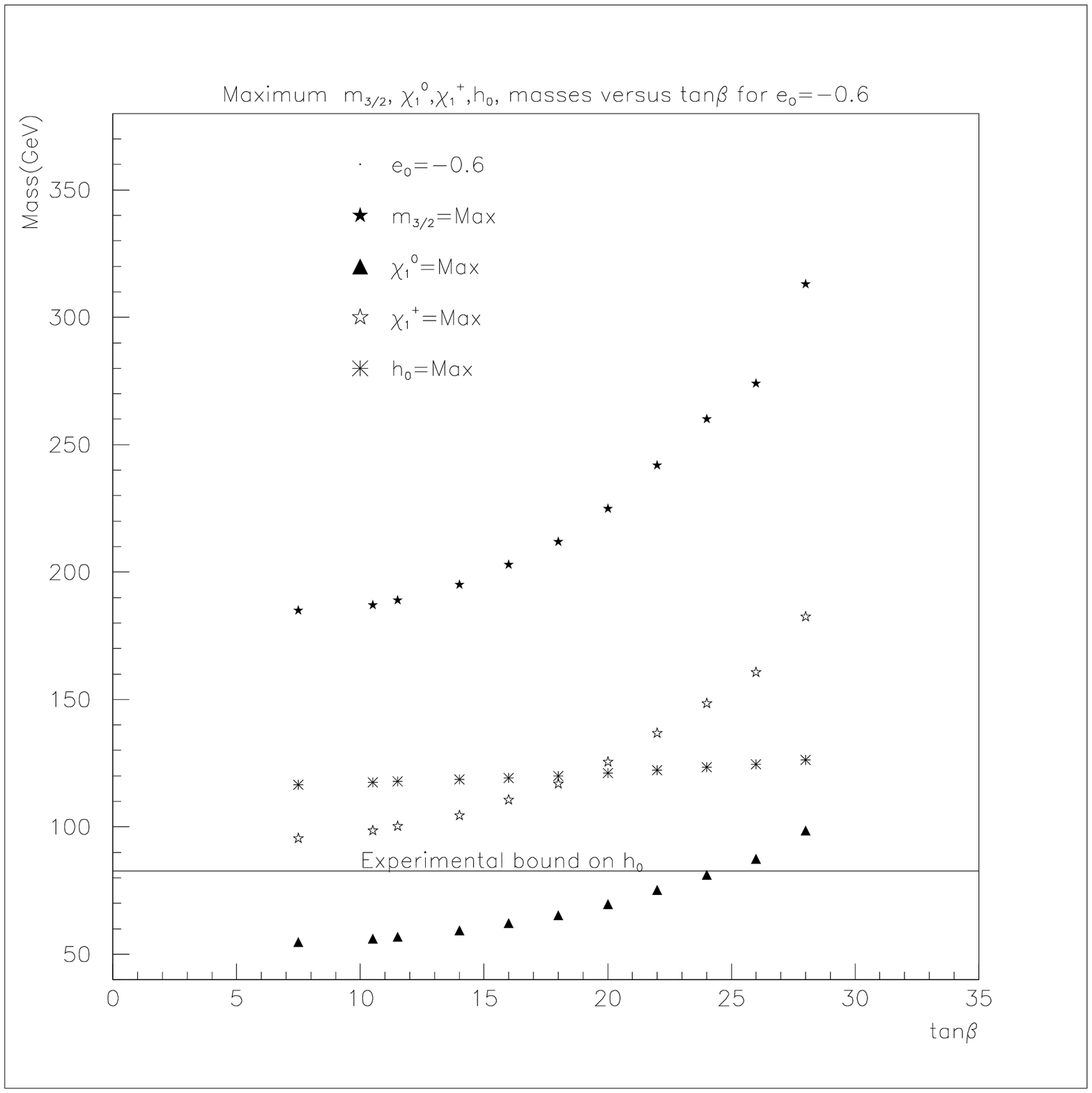}
\caption{Maximum gravitino, lightest chargino and lightest Higgs masses (imposed by 
$\Omega h^2 \leq 0.4$) versus $\tan\beta$ for $e_O=-0.6, \;\;\;\mu>0$.}
\label{grav13}
\end{figure}

In figs.(\ref{fige06},\ref{fige04},\ref{fige041}) we display the 
relic abundance of the lightest 
neutralino  versus 
$\tan\beta$ for different values of the gravitino mass and the 
parameter $e_O$. 
We see that  $\Omega_{LSP} h^2\leq 0.4$ puts $m_{3/2}$ dependent 
lower bounds on the values of $\tan\beta$. 
Let us start the discussion with the case $e_0=-0.6,\mu<0$, fig.(\ref{fige06}).
In this case the upper limit on the relic abundance 
provides the following lower 
bounds  on $\tan\beta$. For instance, for
$m_{3/2}=170$GeV $\tan\beta>5$ while for 
$m_{3/2}=230$GeV $\tan\beta>20$.
Values of the gravitino mass $m_{3/2}<170$GeV come into contradiction 
with the lower experimental bounds on the lightest Higgs mass and lightest 
chargino mass imposed from unsuccessful searches for supersymmetric particles 
in accelerator experiments. 
The lower limit on the relic abundance ($\Omega_{LSP}h^2>0.1$) imposes 
further constraints on the gravitino mass for $\tan\beta\geq 26$. In particularfor $\tan\beta=26$, $m_{3/2}\geq 182GeV$ which results in: 
$m_{\chi_1^{+}}\geq 102GeV,m_{h^0}\geq 115GeV$. Similarly for 
$\tan\beta=28$, $m_{3/2}\geq 210$GeV and 
$m_{\chi_1^{+}}\geq 119.5$GeV. In this case the lightest Higgs mass 
$m_{h^0}\geq 117.6$GeV.
The allowed values of $m_{3/2}$ as a function of $\tan\beta$, 
compatible with (\ref{COSMO}), and for $e_O=-0.6, \mu<0$ are shown in 
fig.\ref{grav}.
Also shown in Figs.\ref{grav13},\ref{grav} are upper bounds on the 
lightest chargino,  lightest 
Higgs and lightest neutralino masses 
respectively,  compatible with  the upper cosmological 
limit on the relic abundance for both signs of $\mu$.
The resulting model can be tested at accelerator experiments.

The cosmological constraints become even more 
important when $e_O \rightarrow 0$. 
In figs. \ref{fige04} and \ref{fige041} we plot the 
relic abundance of the lightest 
neutralino  versus 
$\tan\beta$ for different values of the gravitino mass for $e_O=-0.4$.
The lower experimental bounds (from unsuccessful searches in accelerator 
experiments) on the 
lightest chargino mass and on the 
lightest Higgs mass now require that $m_{3/2}\geq 380 GeV$. 
For $\mu<0$ we see that $\Omega_{LSP} h^2 \leq 0.4$ 
requires $\tan\beta\geq 22$ 
for any allowed value of the gravitino mass and 
for $22\leq \tan\beta \leq 30$, the relic abundance of 
the lightest neutralino is in the range:$0.05\leq \Omega_{LSP} h^2 \leq 0.33$.
The upper bounds on the  gravitino, lightest 
chargino, lightest Higgs, lightest stop and lightest 
stau masses as a function of $\tan\beta$ compatible with the upper 
limit on the relic abundance are summarised in Table 1.
\begin{table}
\begin{center}
\begin{tabular}{|c||c|c|c|c|c|c|}  \hline\hline
$\tan\beta$ &$m_{3/2}^{max}$ & $m_{h^0}$& $m_{\chi_1}^{+}$ & $m_{\chi_1^0}$
& $m_{\tilde{t}_2}$ & $m_{\tilde{\tau}_2}$
\\
\hline\hline
$22 $& 380GeV  & 118.5 GeV& 99 GeV & 54 GeV & 155 GeV  & 310 GeV\\
\hline\hline
$28 $& 381GeV & 118 GeV   & 99 GeV   & 54 GeV &  154 GeV &  284 GeV \\
\hline\hline
$30$ & 400GeV & 119 GeV &  104 GeV  &56 GeV  &  164 GeV  & 288 GeV\\
\hline\hline
$32$ & 495GeV & 122 GeV &  131 GeV & 70 GeV &  216 GeV & 347 GeV \\
\hline\hline
$34$ & 750GeV & 129 GeV &  201 GeV & 106  GeV & 371 GeV & 513. GeV \\
\hline\hline
\end{tabular}
\end{center}
\caption{Upper bounds on  sparticle masses resulting from Eq. ($\ref{COSMO})$ 
for $e_O=-0.4,\mu<0$ for various values of $\tan\beta$.}
\end{table}

\begin{figure}
\epsfxsize=6in
\epsfysize=8.3in
\epsffile{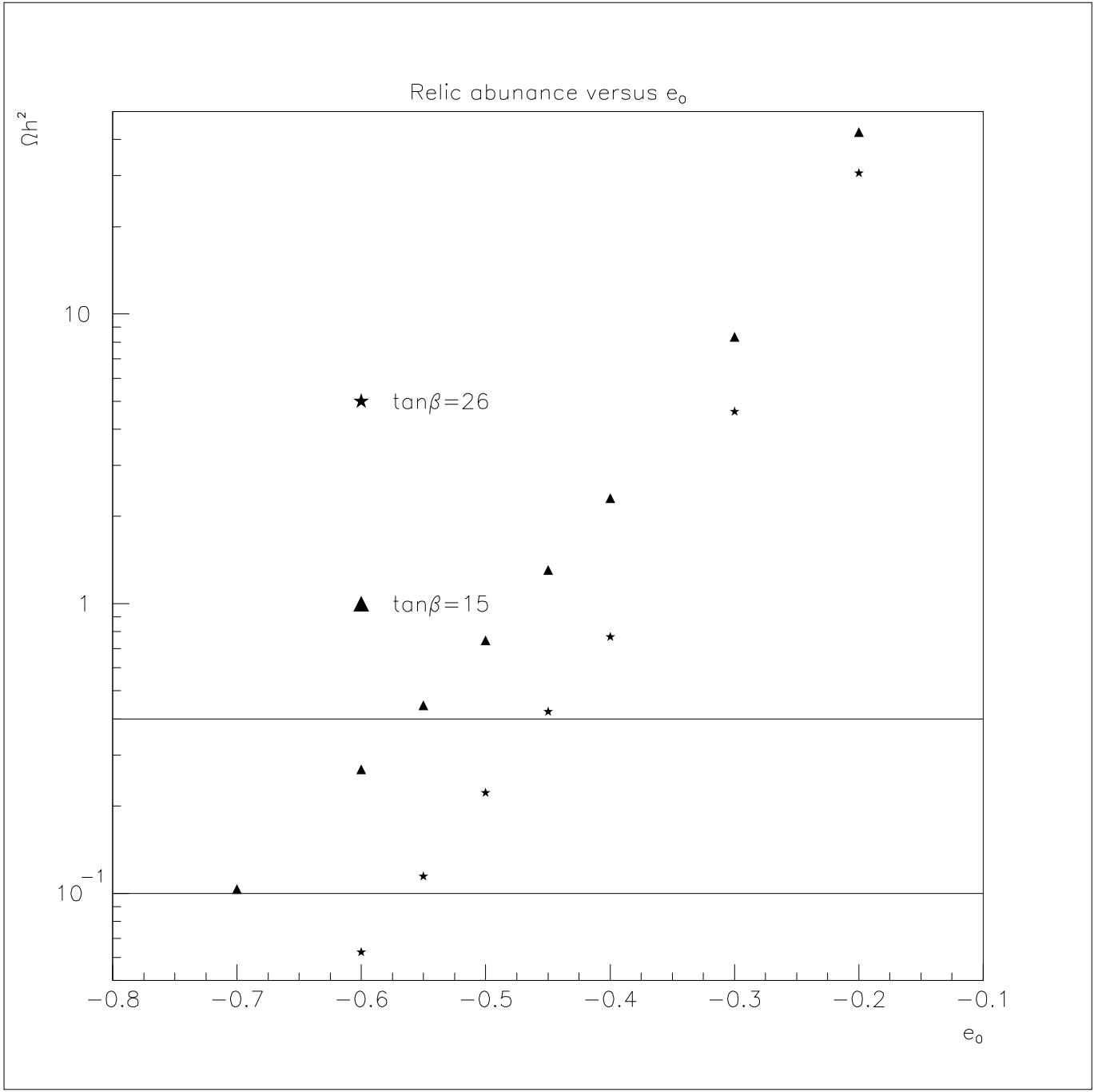}
\caption{Relic abundance of LSP vs $e_O$ for $\tan\beta=15,26,\partial_1 K_5=
\partial_1 \partial_{\bar{1}} K_5=1,|M_1{(M_{U})}|=126GeV,\mu<0.$}
\label{beta15}
\end{figure}

In fig. (\ref{beta15})  we plot 
the relic abundance of the lightest neutralino versus $e_O$ for 
$\tan\beta=15,26$ respectively 
and fixed Bino mass at the unification scale $M_1(M_U)=126$GeV. 
This choice corresponds to a lightest neutralino mass of $\sim 54$GeV
corresponding to a lightest chargino mass of about 100 GeV, the current 
experimental lower bound. 
From fig.\ref{beta15} we observe that for $e_O\geq -0.55$, 
$\tan\beta=15$, $\Omega_{LSP}h^2>0.4$ 
and for $e_O>-0.5$ is greater than 1.  On the other hand for 
$e_O\rightarrow -1$ the 
cosmological constraints are more easily satisfied although for 
$e_O$ too close to -1 $\Omega_{LSP} h^2>0.1$ becomes more difficult to satisfy. Also 
for $e_O<-0.8$ the lightest stau becomes the lightest supersymmetric particle.
For $\tan\beta=26$, we have that for $e_O\geq -0.45$, $\Omega_{LSP}h^2>0.4$.


Assuming that the neutralinos provide the cold dark matter in our Galaxy 
we  calculated its direct detection rates for various nuclei.
For an LSP moving with velocity $v_z$ with respect to the detector nuclei 
the detection rate for a target with mass $m$ is given by \cite{MARK}
\begin{equation}
R=\frac{\rho_{\chi_1^0}^{0.3}}{m_{\chi_1^0}}\frac{m}{A m_p}\int 
f(\bf{v})|v_z|\sigma(|\bf {v}|)d^3 \bf{v},
\end{equation}
where $ \rho_{\chi_1^0}^{0.3}$ denotes the local LSP mass density normalized 
to the standard value of $0.3 {\rm GeV cm^{-3}}$, $ f(\bf{v})$
a Maxwell velocity distribution and $\sigma$ denotes the 
 neutralino-nucleous 
elastic cross section.

For $e_0=-0.6$ the 
cross sections of neutralinos with nucleon 
result in large total detection rates in the cosmologically interesting 
region for large values of $\tan\beta$. 
In particular, for 
$^{73}Ge,^{208}Pb,^{131}Xe$ detectors, 
detection rates of the neutralinos are in the range 
of order $10^{-3}-O(1)$events/Kg/day for $\mu<0$. 
The larger total event rates occur for $\tan\beta\geq 24$.
The results for $\mu>0$ are similar.
This illustrates the fact that  
$\Omega_{LSP}h^2 \sim 
\frac{10^{-37}cm^2}{<\sigma_{anni} v>}$ and the neutralino annihilation 
cross section 
is roughly proportional to the neutralino scattering cross section. 
Thus as the LSP abundance decreases, its scattering cross section 
generally increases.
For $\Omega_{LSP}h^2 \sim 0.1$ this results in an increased event rate.
For values of $e_O>-0.6$ the total detection rates are smaller.
This behaviour of the detection rates can be understood by investigating 
the neutralino-nucleon scalar (spin-independent) cross section, which in 
this model  is the dominant 
contribution to the total neutralino-nucleous 
elastic cross section.

The scalar nucleon-LSP 
cross section is given by \cite{BKL,BOTT}
\begin{equation}
\sigma_{scalar}^{(nucleon)}=\frac{8G_F^2}{\pi}M^2_W m^2_{red}\Biggl[
\frac{G_1(h_0)I_{h_0}}{m^2_{h_0}}+\frac{G_2(H)I_H}{m^2_H}+\cdots\Biggr]^2
\label{scalar}
\end{equation}
where 
\begin{eqnarray}
G_1(h_0)=(-N_{11}\tan\theta_W+N_{21})(N_{31}\sin\alpha+N_{41}\cos\alpha) \nonumber \\
G_2(H)=(-N_{11}\tan\theta_W+N_{21})(-N_{31}\cos\alpha+N_{41}\sin\alpha)
\end{eqnarray}
and 
\begin{equation}
I_{h_0,H}=\sum_{q}l_q^{h_0,H} m_q<N|\bar{q}q|N>
\end{equation}
and 
\begin{eqnarray}
l_q^{h_0}=\frac{\cos\alpha}{\sin\beta}\;\;\; 
l_q^H=\frac{\sin\alpha}{\sin\beta}\;\;{\rm for}\;\;\;\; q=u,c,t \nonumber \\
l_q^{h_0}=-\frac{\sin\alpha}{\cos\beta}\;\;\; 
l_q^H=\frac{\cos\alpha}{\cos\beta}\;\;{\rm for}\;\;\;\; q=d,s,b
\end{eqnarray}
where the two first terms inside the brackets refer to the diagrams 
with $h_0$ and $H$-exchanges in the t-channel and the the ellipsis 
refers to the graphs with squark-exchanges in the s- and u-channels 
\cite{BKL}.
In equation (\ref{scalar}) $m_{red}$ is the 
neutralino-nucleon reduced 
mass, $h_0,H$ denote the lightest Higgs and CP-even heavier Higgs 
respectively and $\alpha$ is the Higgs mixing angle \footnote{We determine 
the Higgs mixing angle numerically by diagonalizing the one-loop CP-even 
Higgs mass matrix.}.

\begin{figure}
\epsfxsize=6in
\epsfysize=8.5in
\epsffile{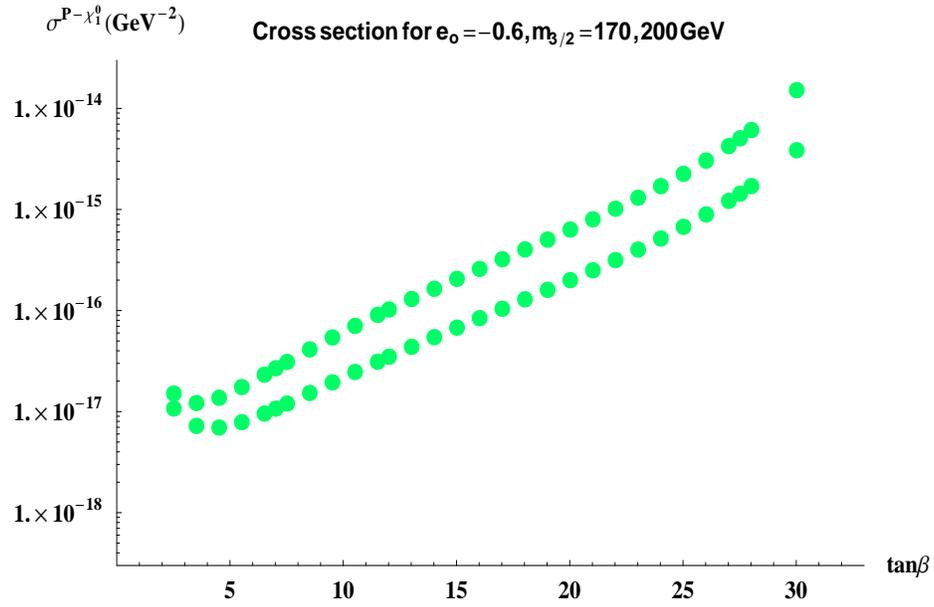}
\caption{Proton-scalar LSP cross section 
versus $\tan\beta$ for fixed $e_o=-0.6,\mu<0, m_{3/2}=170(upper curve),200GeV
(lower curve)$. }
\label{cross2}
\end{figure}

\begin{figure}
\epsfxsize=6in
\epsfysize=8.5in
\epsffile{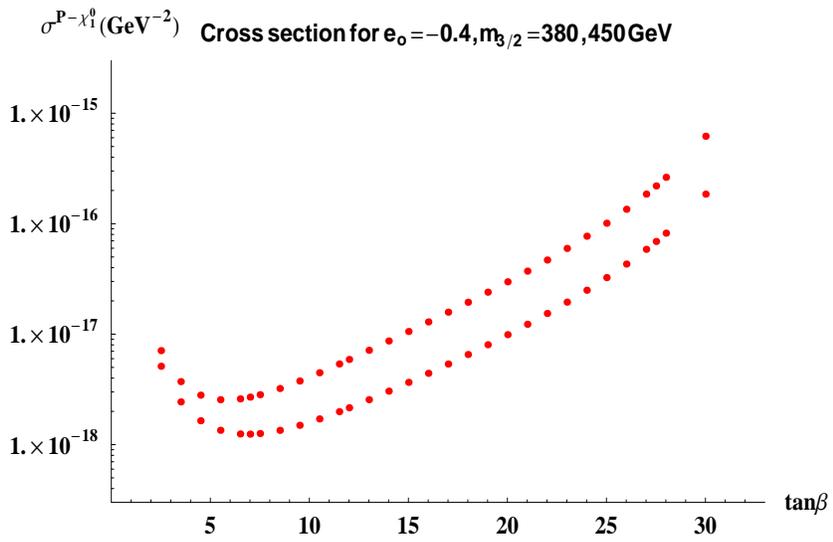}
\caption{Proton-LSP cross section 
versus $\tan\beta$ for  $e_o=-0.4,\mu<0, m_{3/2}=380GeV (upper\;curve),450GeV
(lower\;curve)$. }
\label{cross3}
\end{figure}

Of particular interest is the $\tan\beta$ dependence
of the scalar neutralino-nucleon cross section 
$\sigma_{scalar}^{nucleon}$. For high values of $\tan\beta$ the 
corresponding cross section generically increases (see Figs.\ref{cross2},
\ref{cross3}). The calculated cross section 
for high $\tan\beta$  reaches the sensitivity of current dark 
matter experiments
for the 
nucleon-neutralino scalar 
cross section in the range 
$1\times 10^{-9} {\rm nb} 
(2.6\times 10^{-15}GeV^{-2})\leq \sigma _{scalar}^{(nucleon)} \leq 3\times 10^{-8}{\rm nb}(7.7\times 10^{-14}GeV^{-2})$ \cite{BOTT}.
The larger the value of the $e_O$ parameter the smaller the cross-section 
for fixed $\tan\beta$, sign$\mu$ and fixed bino mass. 
Also the cross section decreases for increasing values of the gravitino mass 
for fixed $e_O$ and sign$\mu$ as is evident from figs.(\ref{cross2},
\ref{cross3}). 
The two upper curves in graphs 
\ref{cross2} and 
\ref{cross3} are cross-sections for an LSP mass about 54GeV. The 
corresponding 
spin-dependent cross sections \cite{BKL} are much smaller by two 
to three orders of 
magnitude.

Recently, 
large detection 
rates have been obtained in type I string 
theories formulated as orientifold compactifications of type IIB 
string theory \cite{CARLOS,Emeis}. In particular, in the mirage unification 
scenario large detection rates have been obtained. This scenario, differs from 
the ones studied in this paper in the following respects:First in the 
mirage unification 
scenario the lightest neutralino has a large Higgsino component while in the 
5-brane dominated limit it is almost a Bino. Thus in the first case 
besides 
a large scalar cross-section the LSP has also a rather large spin-dependent 
couplings with the nuclei. 

Second, in the type I model, the nucleon-neutralino cross 
section is large and consequently the 
detection rates are large when $\tan\beta$ is small, i.e.
$\tan\beta \leq 8$. Also in this case the high neutralino-nucleon 
cross sections correspond to relatively low relic neutralino densities, 
i.e.  $\Omega_{LSP}h^2 \leq 0.1$ and therefore another 
form of dark 
matter might be 
needed to close the 
Universe. As we saw in the current model large cross-sections occur in 
the high $\tan\beta$ region.
Thus the two models lead to different predictions.

As well as predictions for direct detection for the lightest neutralino 
$\chi_1^0$ we have obtained bounds on the value of $\tan\beta$ from the 
cosmological constraints on the relic abundance. We have also calculated 
the maximum lightest chargino mass and the lightest maximum Higgs mass as 
a function of $\tan\beta$ for various values of the ratio of the scalar 
masses to gaugino masses. The resulting sparticle spectra should 
be tested in accelerator experiments in Tevatron and LHC. 

\section*{Acknowledgements}
This research is supported in part by PPARC.

\newpage

\newpage

\newpage

\newpage

\newpage

\newpage

\newpage



\end{document}